\documentclass[useAMS,usenatbib]{mn2e}

\usepackage{graphicx}
\usepackage{rotating}
\usepackage{amsmath}
\usepackage{txfonts}
\usepackage{color}
\usepackage{ulem}
\usepackage[hyphens]{url}
\usepackage{hyperref}

\title[High-precision astrometry with VVV. I.]{High-precision
  astrometry with VVV. I. An independent reduction pipeline for
  VIRCAM@VISTA\thanks{Based on observations with the 4\,m VISTA ESO
    telescope.}}

\author[Libralato et al.]{M.\ Libralato\thanks{E-mail:
    \href{mailto:mattia.libralato@studenti.unipd.it}{mattia.libralato@studenti.unipd.it}}\thanks{Visiting
    Ph.D. Student at STScI under the 2013 DDRF Spring
    program.}$^{1,2,3}$, A.\ Bellini$^{3}$, L.\ R.\ Bedin$^{2}$,
  J.\ Anderson$^{3}$, G.\ Piotto$^{1,2}$, V.\ Nascimbeni$^{1,2}$,
  \newauthor I.\ Platais$^{4}$, D. Minniti$^{5,6,7}$,
  M. Zoccali$^{7,8}$ \\
$^{1}$ Dipartimento\ di Fisica e Astronomia, Universit\`a di Padova,
  Vicolo dell'Osservatorio 3, Padova, I-35122, Italy \\ $^{2}$
  INAF-Osservatorio Astronomico di Padova, Vicolo dell'Osservatorio 5,
  Padova, I-35122, Italy \\ $^{3}$ Space Telescope Science Institute,
  3700 San Martin Drive, Baltimore, MD-21218, USA \\ $^{4}$ Department
  of Physics and Astronomy, The Johns Hopkins University, Baltimore,
  MD-21218, USA \\ $^{5}$ Departamento de Ciencias Fisicas,
  Universidad Andres Bello, Republica 220, Santiago, Chile \\ $^{6}$
  Vatican Observatory, V00120 Vatican City State, Italy \\ $^{7}$
  Millenium Institute of Astrophysics, Av. Vicu\~na Mackenna 4680,
  Macul, Santiago, Chile \\ $^{8}$ Instituto de Astrof\'isica,
  Facultad de F\'isica, Pontificia Universidad Cat\'olica de Chile,
  Av. Vicu\~na Mackenna, Santiago, Chile \\ }

\begin{document}

\date{Received 03 December 2014 / Accepted 25 March 2015}

\pagerange{\pageref{firstpage}--\pageref{lastpage}} \pubyear{2015}

\maketitle

\label{firstpage}

\begin{abstract}
We present a new reduction pipeline for the \mbox{VIRCAM@VISTA}
detector and describe the method developed to obtain high-precision
astrometry with the VISTA Variables in the V\'ia L\'actea (VVV) data
set. We derive an accurate geometric-distortion correction using as
calibration field the globular cluster NGC~5139, and showed that we
are able to reach a relative astrometric precision of about 8 mas per
coordinate per exposure for well-measured stars over a field of view
of more than 1 square degree. This geometric-distortion correction is
made available to the community. As a test bed, we chose a field
centered around the globular cluster NGC~6656 from the VVV archive and
computed proper motions for the stars within. With 45 epochs spread
over four years, we show that we are able to achieve a precision of
1.4 mas yr$^{-1}$ and to isolate each population observed in the field
(cluster, Bulge and Disk) using proper motions. We used
proper-motion-selected field stars to measure the motion difference
between Galactic disk and bulge stars. Our proper-motion measurements
are consistent with UCAC4 and PPMXL, though our errors are much
smaller. Models have still difficulties in reproducing the
observations in this highly-reddened Galactic regions.
\end{abstract}


\section{Introduction}
\label{intro}

The VISTA Variables in the V\'ia L\'actea (VVV) variability campaign
started in 2010. Thanks to the VISTA InfraRed Camera (VIRCAM,
\citealt{Dal06,Eme06}), mounted at the 4.1\,m telescope VISTA (Visible
and Infrared Survey Telescope for Astronomy), this ongoing survey is
mapping the Galactic bulge and disk to create a 3-D map of the Milky
Way \citep{Min10,Sai12}. As for many long-term variability surveys,
the observing strategy is mainly focused on covering a portion of the
sky as large as possible in a single night, scanning the full field of
view many times every few days. To this aim, the exposure time of each
image has to be short enough in order to achieve the survey
specifications. In the VVV survey, the typical exposure time for
$K_{\rm S}$-filter images is about 4\,s, e.g., a factor 7 smaller than
the 30-s threshold set by \citet{Pla02} and \citet{Pla06} as the
minimum exposure time required to average out the large-scale
semi-periodic and correlated atmospheric noise that harms ground-based
astrometry. In spite of this, we chose to exploit the astrometric
capabilities of this survey that will release to the community a data
set with more than one hundred epochs over six years.

In this paper, we present our reduction pipeline for the \mbox{VIRCAM}
detectors and the geometric-distortion correction. As an example, we
also show a few applications made possible by the astrometric accuracy
reached by the VVV data set so far.

\begin{table}
  \caption{List of the VIRCAM@VISTA data used for the astrometric
    calibration. Each observing block is made by $N_{\rm step}$
    images, where ``step'' is the dither spacing in arcmin between two
    consecutive exposures in an observing block. The single-image
    exposure time is given by the integration time (DIT) multiplied by
    the total number of individual integrations (NDIT).}  \centering
  \label{tab:obs}
  \begin{tabular}{ccccc}          
    \hline\hline
    \textbf{Filter} & \textbf{$N_{\rm step}$} & \textbf{Exposure Time} & \textbf{Seeing} & \textbf{Airmass} \\ 
    & & (NDIT$\times$DIT) & (arcsec) & ($\sec z$) \\
    \hline  
    & \\
    \multicolumn{5}{c}{\textbf{Program ID: 488.L-0500(A) -- PI: Bellini}} \\
    \multicolumn{5}{c}{{\bf NGC~5139 -- $\omega$ Cen}} \\
    & \\
    \multicolumn{5}{c}{\it February 23-24, 2012} \\
    & \\
    $J$ & $25_{1.2}$ & (6$\times$10\,s) & $0.97$-$1.42$ & 1.026-1.107 \\
    $J$ & $25_{8.5}$ & (6$\times$10\,s) & $0.74$-$1.08$ & 1.134-1.198 \\
    & \\
    \hline
  \end{tabular}
\end{table}

\section{Instrument and observations}
\label{OBS}

VIRCAM is a mosaic of 4$\times$4 detectors mounted at the focus of the
VISTA 4.1\,m telescope. Each detector is a Raytheon VIRGO
2048$\times$2048-pixel array and covers $\sim$694$\times$694
arcsec$^2$ on the sky.  The average pixel scale is
$0^{\prime\prime}\!\!.339$ pixel$^{-1}$ \citep{Suth14}. The gaps
between the detectors are quite large and correspond to 42.5\% and
90\% of the detector size along the $X$ and $Y$ direction,
respectively. 

Dithered observations are recommended to self-calibrate the geometric
distortion of a detector (e.g., \citealt{Ande06}, \citealt{BB10},
\citealt{Libra14}). However, the standard dither pattern adopted by
VVV is not adequate for this purpose. For this reason, a calibration
program (Program ID: 488.L-0500(A), PI: Bellini) was approved in
2012. The calibration field is centered on globular cluster NGC~5139
($\omega$\,Cen). This field was chosen due to its
\mbox{relatively-high} star density over more than one square degree.

\begin{figure}
  \centering
  \includegraphics[width=9.cm]{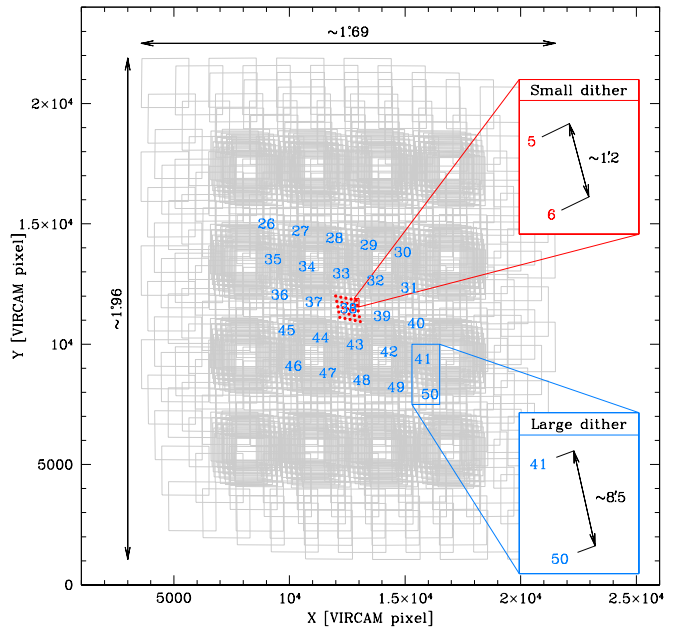}
  \caption{Outline of our adopted dither pattern used for $J$-filter
    data. Large- and small-dither images are organized in two
    5$\times$5 arrays. The pointings of the large-dither images are
    taken to cover the gap between the 16 detectors. The total field
    of view covered by our observations is about
    $1^\circ\!\!.7$$\times$$2^\circ\!\!.0$ on sky. The two zoomed-in
    panels show the dither spacing for large (azure panel) and small
    (red panel) dither pattern.}
  \label{fig1}
\end{figure}

The field was observed in the $J$ filter in two runs of 25 images each
(Table~\ref{tab:obs}), both organized in an array of 5$\times$5
pointings, but with a different dither spacing
(Fig.~\ref{fig1}). Large dithers were taken to cover the gap between
the 16 chips and to allow us to construct a single reference system
for all observations. The small dither pattern was obtained to allow
independent modeling of the \mbox{high-frequency} residuals of the
geometric distortion for each chip.

\section{Data reduction}
\label{datared}

We developed a reduction package that makes use of the same tools
described in \cite{Libra14} for the HAWK-I detector. Here, we briefly
describe the software, and focus on the few differences between the
two works. \\

One peculiarity of VIRCAM is the striped pattern that affects all the
images, both calibration and scientific. These stripes are generated
by the IRACE electronics \citep{Suth14} and change from one exposure
to the next. To correct them, we made a FORTRAN routine based on the
Cambridge Astronomy Survey Unit (CASU) pipeline correction, that
resembles the correction applied by \cite{May08} for the
WFPC2@\textit{HST} background streaks. We computed in each image the
clipped-median value of the counts in each row, then we took the
median of these values and subtracted it from the clipped-median value
of each row. These differences represent the corrections to be applied
to each row. We did not include bad/warm/hot pixels while computing
the median values.

Using archival flat-field images we were not able to completely
correct for the pixel-to-pixel sensitivity variation of the detectors,
in particular for chip [16] where the very high quantum-efficiency
variation on short timescales sometimes makes it impossible to
properly apply the flat-field
correction\footnote{\href{http://casu.ast.cam.ac.uk/surveys-projects/vista/technical/known-issues}{http://casu.ast.cam.ac.uk/surveys-projects/vista/technical/known-issues}}. So,
we constructed master flat-field frames using the scientific images
themselves, masking all bad/warm/hot pixels and those in close
proximity of any significant source (stars and galaxies) and
considered these purged images as on-sky flat-field images.

First, we applied dark and flat-field corrections to all images. Then,
for each chip we computed the median sky value in a 5$\times$5 grid
and subtracted it according to the table. Then we made a 5$\times$5
grid of fully-empirical PSF models for each detector of each exposure,
following the prescription given in \cite{Ande06}. Unlike the
procedure given in the original {\sf{\mbox{img2psf\_WFI}}} and
{\sf{\mbox{img2psf\_HAWK-I}}} programs, the finding criteria (minimum
flux and minimum separation from brighter stars) to choose the stars
that would be used to model the PSF are applied locally and are
different in each cell of the grid. This way, we are able to find the
most suitable combination of these criteria in each of the 5$\times$5
regions of the chip (e.g., if the cluster center is located in a
corner of a chip, the minimum separation from the brighter stars in
that corner is usually lower than in the opposite corner where the
crowding is lower). With an array of PSF models, we are able to
measure positions and fluxes for all sources on an image. The final
catalogs (one for each chip) contain positions, instrumental
magnitudes\footnote{Defined as $-2.5\times \log (\sum \rm{counts})$,
  where $\sum\rm{counts}$ is the sum of the total counts under the
  fitted PSF.}, and another quantity called quality-of-PSF-fit
({\sf{QFIT}}) which represents the absolute fractional error in the
PSF-model fit to the star \citep{Ande08}. The lower the {\sf{QFIT}},
the better is the PSF fit. The {\sf{QFIT}} parameter is a useful
quantity to discriminate among well-measured and poorly-measured
stars. Typically, in $\omega$~Cen catalogs we considered bright,
unsaturated stars with a {\sf{QFIT}}$<$0.05 to be well-measured
stars. These selections allow us to always have at disposal over 100
stars per chip, with an average value of 350 well-measured stars for a
corner chip and 1000 stars for a centermost chip.

\begin{figure*}
  \centering
  \includegraphics[width=\textwidth]{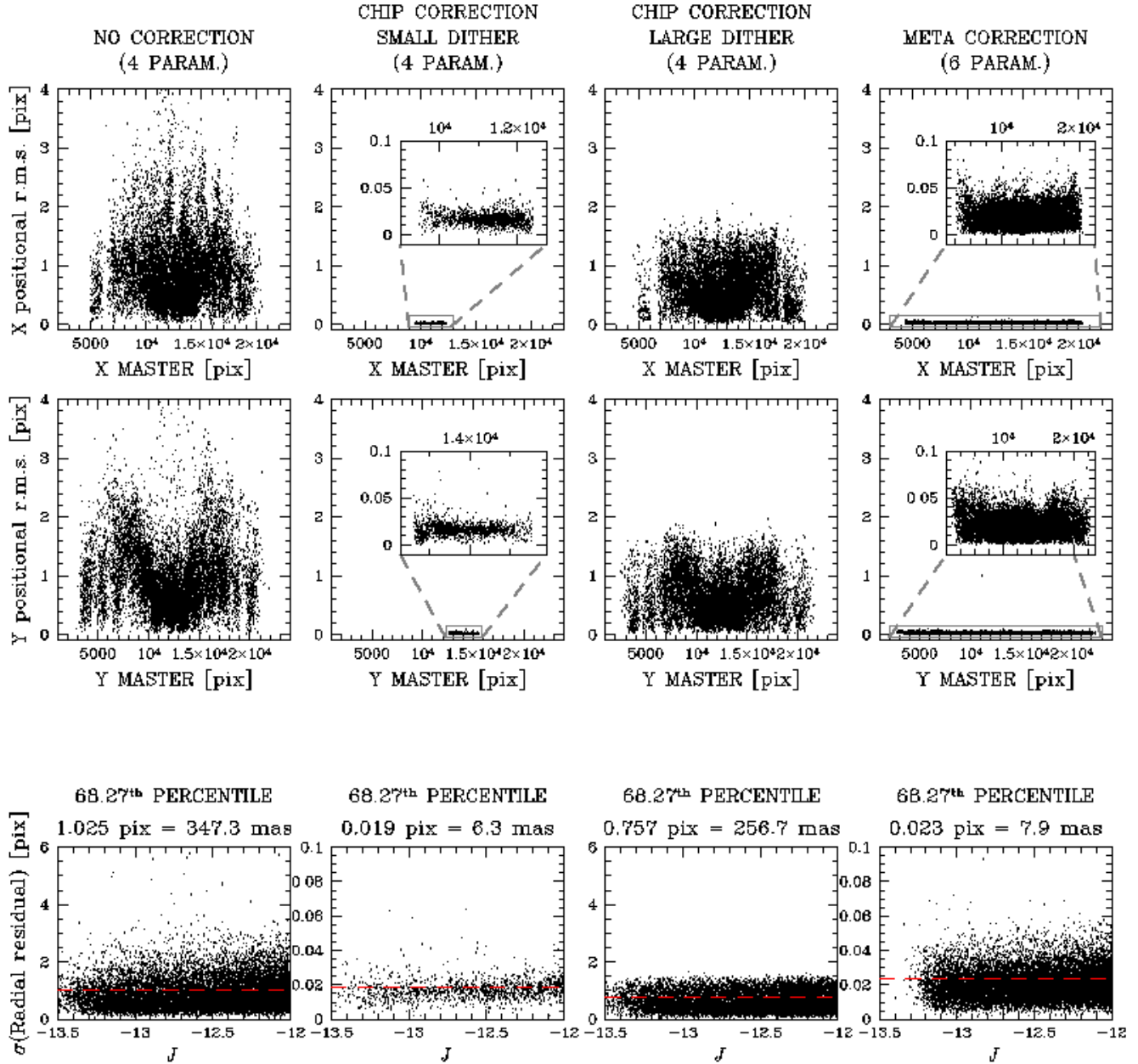}
  \caption{In each column of the Figure we show the positional
    residuals for different master frames. (\textit{Left}): master
    frame based on single-chip, uncorrected catalogs and conformal
    transformations. (\textit{Middle-left}): chip \#10 master frame
    based on single-chip correction, small dithers, and conformal
    transformations. (\textit{Middle-right}): master frame based on
    single-chip correction, large dithers, and conformal
    transformations. Even if there is an improvement with respect to
    the ``\texttt{NO CORRECTION}'' case, the bumps (caused by
    projection-induced effects) are still visible. (\textit{Right}):
    master frame based on our final distortion solution, six-parameter
    linear transformations and taking into account the projection
    effects. In the \textit{Top}(\textit{Middle}) row, we show the
    positional r.m.s. along the $X$($Y$) axis as a function of the
    $X$($Y$) position on the master frame. In the \textit{Middle-left}
    and \textit{Middle-right}-panel insets, we zoomed-in to better
    show the residuals. In the \textit{Bottom} panels we plot the
    $\sigma$(Radial residuals) as a function of the $J$ magnitude. The
    red dashed lines are set to the 3$\sigma$-clipped 68.27$^{\rm
      th}$-percentile value of $\sigma$(Radial residuals). In all
    these panels we show only bright, unsaturated and well-measured
    stars. See text for more details.}
  \label{fig2}
\end{figure*}

\section{Geometric-distortion correction}
\label{GDC}

In the large field of view (FoV) of VIRCAM, the tangential-plane
projection effects are not negligible (at one degree from the tangent
point this corresponds to more than 0.18 arcsec, $\sim$0.5 VIRCAM
pixel). This means the farther from the center, the larger is the
difference between the true position and the projected position of a
star. 

We chose to perform an auto-calibration. By using as a reference
system 2MASS \citep{Skru06} or UCAC4 \citep{Zac13}, which are among
the most accurate absolute systems, we would have unavoidably ended up
limited by their accuracy (of the order of 0.2-0.3 arcsec for
2MASS). Not to mention the non-negligible contribution from the
stellar motion between the reference system and our
exposures. Furthermore, as stated in \cite{BB10}, it is difficult to
find a distortion-free reference frame with an homogeneous stellar
density and luminosity. Therefore we adopted the auto-calibration
solution. The basis of the auto-calibration is to observe the same
star in as many different locations on the detector as possible and to
compute its average position once it is transformed onto a common
reference frame. Thanks to the large number and varied spacing of our
dither pattern, a given star will be observed in several different
locations in the FoV and, as such, the systematic errors in its mean
position should average out. This way, the average positions of the
stars should provide a reasonable approximation of their true
positions in a distortion-free frame (the master frame). We built the
master frame by cross-identifying the stars in each single-detector
catalog of each exposure. We used conformal transformations
(four-parameter linear transformations: rigid shifts in the two
coordinates, one rotation, and one change of scale) to bring the
stellar positions measured in each image into the reference system of
the master frame. In the left panels of Fig.~\ref{fig2} we show the
effects of the projection on the master frame. The positional
residuals along the $X$ and $Y$ axes show several bumps where two
different chips overlap.

When we first examined plots like these, it was clear that the bumps
at the boundaries of the chips could be due either to internal
distortions within each chip or to errors in placing the chips
properly with respect to each other. To ensure that the distortion
within each chip was properly accounted for, we independently solve
for the geometric distortion of each chip, as done in
\cite{Libra14}. For this specific purpose, we only used the
small-dither images where the gaps between the chips are not covered
(no detector overlaps with any other detector). The accuracy of the
single-chip distortion solution was at the 0.02-pixel level ($\sim$7
mas). As an example, in the middle-left panels of Fig.~\ref{fig2} we
show the residuals after we applied the single-chip correction to the
catalogs and constructed a small-dither-based master frame using only
chip \#10. As shown in Fig.~\ref{fig1}, the small dithers do not allow
us to put all the 16 chips in the same reference system since the gaps
are so large that the chips do not overlap each other. For this reason
we selected one random chip (chip \#10) to show that our single-chip
distortion solution was good as we wrote above. Then, we applied our
single-chip correction to all catalogs and used four-parameter linear
transformations to create a new master frame based on large-dither
images. Again, the positional-residual bumps were still visible
(middle-right panels of Fig.~\ref{fig2}). Therefore, these trends in
the positional residuals are ascribable to projection effects, which
have to be taken into account while cross-identifying the catalogs.

We chose to define a meta reference system in which to properly
project all single-chip catalogs and, at the same time, solve for most
of the geometric distortion that affects this detector. We proceeded
as follows. We used the 2MASS catalog as our initial reference
frame. We projected the 2MASS catalog onto a tangent plane centered on
$\omega$\,Cen and followed the prescriptions given in \cite{Van06} to
convert R.A. and Dec. positions into pixel-based coordinates. This is
an important step because we imposed the master-frame scale to be
exactly equal to 0.339 arcsec pixel$^{-1}$ for all chips. This value
is the average pixel scale declared by \cite{Suth14}.

Initially, we cross-identified all stars of each single-chip raw
frames with the 2MASS catalog by using six-parameter linear
transformations (which also include the deviation from orthogonality
and the change of relative scale between the two axes). Then, we
located the center of each chip ($x$,$y$)$=$(1024,1024) on the
2MASS-based reference system. Without properly taking into account for
the projection effects, the chip-center positions on the 2MASS
reference frame depend on their distances from the tangent point
($\omega$\,Cen center) and on the geometric distortion. To get rid of
the first dependency and find the best position of the chip centers,
we iteratively de-projected the 2MASS catalog onto the celestial
sphere, and then projected it again using as the tangent point the
current chip-center position on the 2MASS reference system, in order
to compute new, improved transformations. For each chip of each
exposure/image we iterated the whole process five times (after the
fifth iteration the adjustments were negligible).

Once the chip centers in the 2MASS reference frame converged to fixed
positions, in order to build the meta-reference system, we had to
impose additional constraints. First, the meta center was defined as
the average position of the four centermost chips. The second
constraint we imposed is that the $Y$ and $X$ axes of our meta-frame
system had to be oriented up and to the right, respectively. For each
of the four centermost chips, we computed the angle between the
expected meta-frame $X$ axis and the segment that connects the center
of the meta frame to the chip center. Then we rotated all chips by the
average of the four angles.

For each image we de-projected the 2MASS frame onto the celestial
sphere and projected it back on a tangent plane, but this time using
as tangent point the meta center computed as described above. Then, we
rotated and shifted these 2MASS-based positions according to the other
constraint. The final products of this effort are 2MASS-based
positions projected on the meta-frame center of each image, rotated
and shifted to have the meta center in ($x$,$y$)$=$(0,0). These
positions represent the best approximation of the expected
distortion-free meta positions.

For each star in common between our catalogs and the 2MASS-based
catalog, we have a pair of positional residuals that correspond to the
difference between the raw-chip positions and the expected meta-frame
positions (given by the stellar positions on the modified 2MASS
reference frame). We used both saturated and unsaturated stars with
magnitude $J$$<$-12 and {\sf{QFIT}}$<$0.2. Since 2MASS is a shallow
survey, we had to use saturated ($J$$\lesssim$$-$13.4) stars in order
to have an adequate sample size. We divided each chip into a
3$\times$3-grid elements and, in each such element, we defined the
grid-point value as the average value of the residuals within. The
cells have different sizes, with those close to the edges (for example
512$\times$512 pixels on the corners) smaller than the central one
(1024$\times$1024 pixels), in order to better model the distortion
close to the edges. As described in \cite{Libra14}, for those cells
adjoining the detector edges we shifted the grid points to the
edge. We built a look-up table of correction for any location of the
chip, using a bi-quadratic interpolation among the surrounding four
grid points. To avoid extrapolation, our grid points extended to the
corners, but this meant that we needed several iterations (each time
applying 90\% of the suggested correction to the raw positions and
computing new residuals) before convergence could be achieved. \\

After this first part of the correction, we have star positions
transformed into a meta reference frame and corrected for geometric
distortion. All the stellar positions collected in one meta catalog
are those of the stars imaged in one exposure. Therefore, accordingly
to the Table~\ref{tab:obs}, we have 50 meta catalogs at our disposal
(25 of which are based on large-dither exposures, while the other 25
are based on the small-dither exposures).  The astrometric accuracy
achieved is about 0.2--0.3 pixel, similar to that of 2MASS. The
astrometric quality of our measurements should be ten times better
than this, so to further improve our result, we applied an additional
table-of-residuals correction to each chip by comparing the positions
of the stars as measured in different meta catalogs, thus enabling a
precision of $\sim$0.03 pixel per comparison, as follows. For each
pair of meta catalogs (hereafter catalogs \#1 and \#2), we
cross-identified all stars in common by using six-parameter linear
transformations. We found the meta center of catalog \#1 into the
reference system of catalog \#2 and projected the stellar positions
measured in catalog \#2 into the tangent plane centered at the center
of catalog \#1. Then we computed the positional residuals as the
difference between the stellar positions in the meta \#1 reference
system and the positions in the meta \#2 reference system, once
projected and transformed into the meta \#1 reference system. For
those meta catalogs obtained from the large-dither images, we compared
each of them to the other 49 catalogs, while small-dither catalogs
were only compared to the large-dither ones. When we computed the
distortion correction for each chip individually, we used only
small-dither images. We then applied this correction to large-dither
images and looked at the residuals computed by comparing our stellar
positions with those of 2MASS. We noticed that the non-linear terms of
the distortion over a very large scale were not completely accounted
for. Therefore, we chose to compute the positional residuals by
comparing only images far enough on the sky from each other. For each
chip, we collected all these residuals together and divided them into
an array of 11$\times$11 square elements. We assigned to each array
element the median value of the residuals within. For any location on
the chip, the correction is computed as the bi-linear interpolation
between the surroundings four grid points. We iterated five times,
computing new residuals and adding the new correction to the previous
one. 

In summary, the distortion solution of each chip consists of two
parts: a 3$\times$3 look-up table of residuals (that is used to
compute the correction at any inter-chip location via a bi-quadratic
interpolation between the surrounding four grid points), and an
additional fine-tuning 11$\times$11 look-up table of residuals (this
time using a bi-linear interpolation to compute the correction). The
final stellar positions are distortion corrected and projected with
respect to the center of the meta catalog. Therefore, each meta
catalog is projected into a different tangent plane. It is important
to transform all the catalogs into the same tangent plane during the
construction of the master frame. In the rightmost panels of
Fig.~\ref{fig2} we show the result of our efforts. We applied our
distortion correction to the stars in each meta catalog. We used
six-parameter linear transformations to bring these corrected
positions on the master-frame reference system using the same tangent
plane for each catalog. This way, the $\sigma$(Radial
residuals)\footnote{The $\sigma$(Radial residuals) is defined as:
\begin{displaymath}
  \sigma(\textrm{Radial residual})_{i} = \sqrt{\frac{ (x_{i,j}^{T_{j}}
      - X_{i}^{\rm master})^2 + (y_{i,j}^{T_{j}} - Y_{i}^{\rm
        master})^2 }{ 2 }}\phantom{1} ,
\end{displaymath}
where ($x_{i,j}^{T_{j}},y_{i,j}^{T_{j}}$) is the position of the
$i$-th star of the $j$-th image, distortion corrected and transformed
into the master-frame reference system, $T_j$ is the the
transformation of the $j$-th image into the master frame, and
($X_{i}^{\rm master},Y_{i}^{\rm master}$) is the distortion-free
(master-frame) position of the $i$-th star.} improves from $\sim$1.025
pixels (347.3 mas) to 0.023 pixel (7.9 mas).

In Appendix~\ref{appendix} we show the distortion maps and the
positional residuals along the $X$ and $Y$ axes, before and after the
correction, for each of the 16 chips of VIRCAM. \\

With this paper, we release a \texttt{FORTRAN} routine to correct the
geometric distortion. It requires the single-chip raw coordinates
($x^{\rm raw}$,$y^{\rm raw}$) and the chip number. In output, the code
computes ($x^{\rm corr}$,$y^{\rm corr}$) coordinates in the meta-frame
reference system. The code is available at our group's web
page\footnote{\href{http://groups.dfa.unipd.it/ESPG/}{http://groups.dfa.unipd.it/ESPG/}}.

\section{Application: NGC~6656}

To test our geometric-distortion correction we computed relative
proper motions (PMs) of stars in the field of the globular cluster
NGC~6656, $(\alpha,\delta)_{\rm J2000.0}=(\rm
18^h36^m23^s\!\!.94,-23^{\circ}54^{\prime}17^{\prime\prime}\!\!.1$,
\citealt{Har96}, 2010 edition), using VVV data. We chose this object
for its closeness and relatively-high PM with respect to the field
objects. We used images taken between 2010 and 2014. We have 12 images
in each of the 45 epochs used (except for one epoch for which we have
14 images) in the $K_{\rm S}$ filter (from 2010 to 2014), while for
the $J$ filter there are only 12 images taken in 2010.

\begin{figure}
  \centering
  \includegraphics[width=\columnwidth]{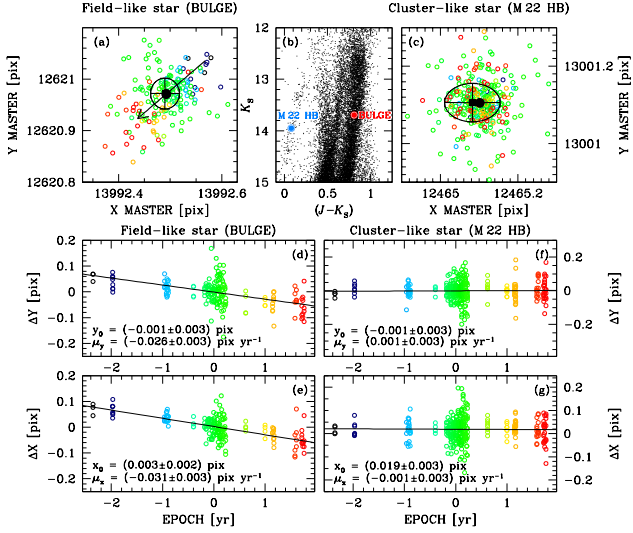}
  \caption{Multi-epoch fit of PMs for a field-like (``BULGE'') and a
    cluster-like (``M\,22 HB'') star. (\textit{Top}): in panels (a)
    and (c), the empty circles represent star's positions at different
    epochs transformed into the reference master frame,
    \mbox{color-coded} (as defined in the bottom panels) depending on
    the time interval relative to the reference epoch
    (J2012.62423). The master-frame position is represented by the
    solid black square (surrounded by an ellipse with semi-axes equal
    to the positional r.m.s. along the $X$ and $Y$ direction); the
    solid black circle is the expected position of the star at the
    reference epoch based on the PM fit. In the panel (a), the black
    arrow shows the $\sim$4-yr displacement of the star. In panel (b)
    we show the position of the selected stars on the
    CMD. (\textit{Bottom}): Motion in $Y$ [panels (d) and (f)]
    and in $X$ [panels (e) and (g)] as a function of
    the time from the reference epoch. The black line is the
    least-squares fit of the PM.}
  \label{fig3}
\end{figure}

\begin{figure}
  \centering
  \includegraphics[width=\columnwidth]{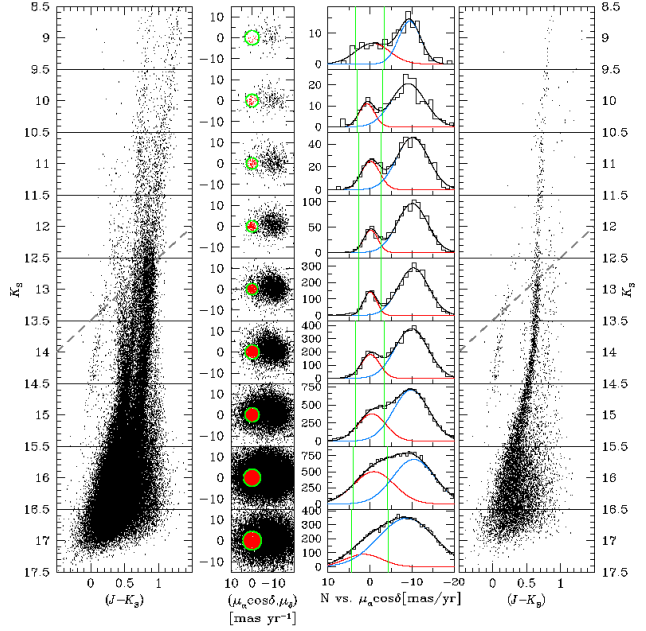}
  \caption{(\textit{Left}): $K_{\rm S}$ vs. ($J$$-$$K_{\rm S}$) CMD of
    the NGC~6656 field. We show only well-measured-PM stars. We split
    the CMD in nine intervals of one mag each. The gray dashed line
    sets the average saturation threshold. The saturation level
    slightly varies within each VIRCAM chip. The variation becomes
    substantial across the total FoV, and of course from one exposure
    to the next (because of generally different seeing
    conditions). (\textit{Middle-left}): VPDs for each of the
    corresponding magnitude interval. The mean motion of cluster
    members is centered at (0,0) in the VPDs. We plotted with red dots
    the \mbox{cluster-like} stars. The radius for the cluster-member
    selection (green circle) ranges from $\sim$4.4 mas for stars with
    16.5$<$$K_{\rm S}$$\le$17.5 to $\sim$4.1 mas for those stars with
    8.5$<$$K_{\rm S}$$\le$9.5. (\textit{Middle-right}): Histograms for
    the $\mu_\alpha\cos\delta$ proper motion distribution. The bin
    size changes depending on the total number of stars in each
    magnitude bin. Dual-Gaussian fit in black; individual Gaussians in
    red and azure are used for cluster and field
    $\mu_\alpha\cos\delta$ distributions respectively. The field
    distribution is wider than that of the cluster and contaminate the
    cluster-member sample in all magnitude bins. (\textit{Right}): CMD
    with only cluster-like-motion stars. It is clear that the fainter
    the magnitude bin, the higher is the field contamination in our
    sample. The PMs have been corrected for differential-chromatic
    refraction as described in the text.}
  \label{fig4}
\end{figure}

\begin{figure}
  \centering
  \includegraphics[width=\columnwidth]{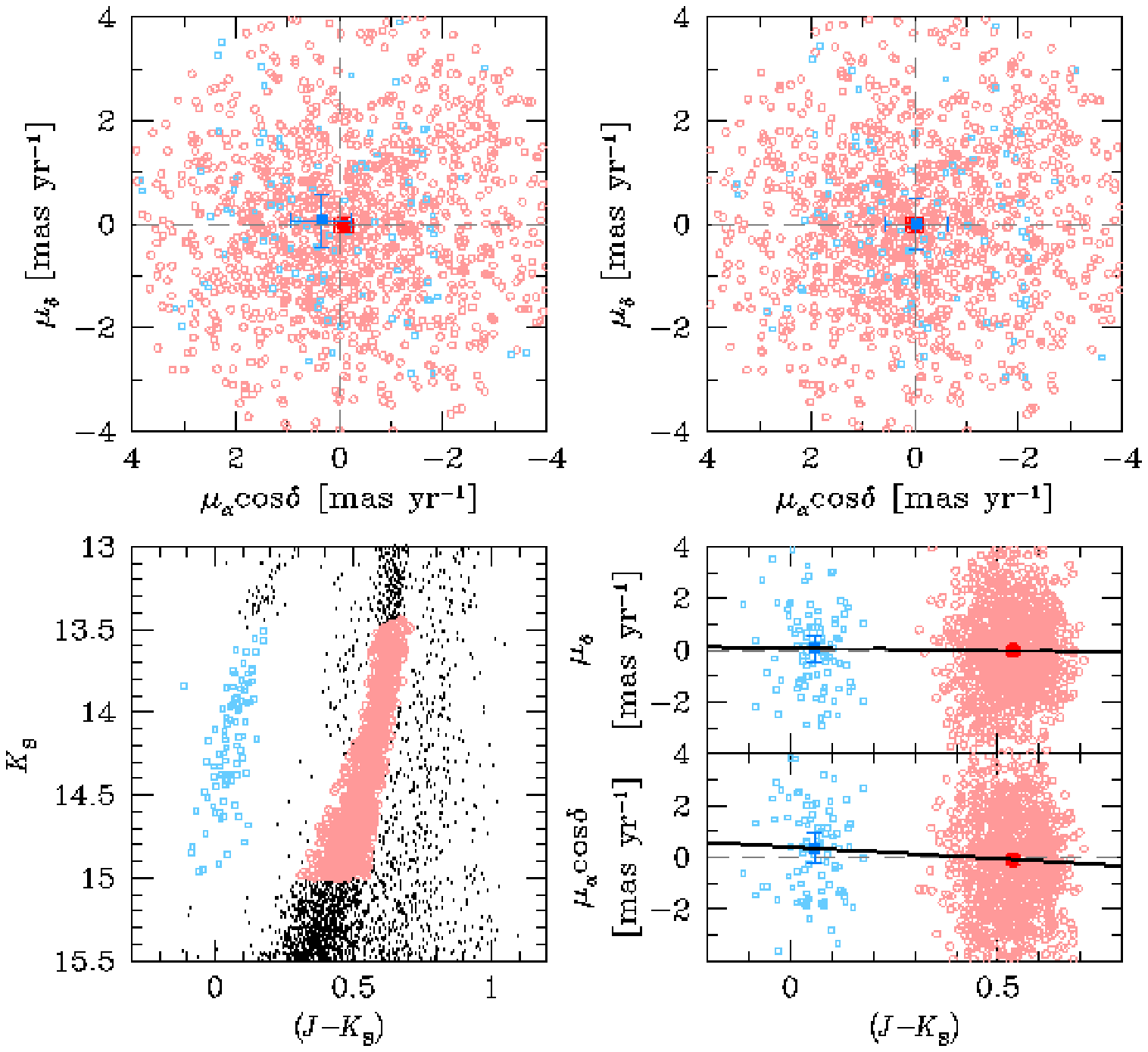}
  \caption{(\textit{Bottom-left}): $K_{\rm S}$ vs. ($J$$-$$K_{\rm S}$)
    CMD of NGC~6656. We plotted the horizontal-branch stars with light
    blue open squares and red-giant-branch stars with light red open
    circles. We considered only stars with $K_{\rm S}$ between 13.5
    and 15. (\textit{Top-left}): VPD with only the blue and red stars
    before the differential-chromatic-refraction correction. The azure
    solid square represents the mean motion of the blue stars, while
    the red solid circle that of the red stars. We also plot the
    3$\sigma$ error bar as reference. (\textit{Bottom-right}):
    $\mu_\alpha\cos\delta$ and $\mu_\delta$ as function of the
    ($J$$-$$K_{\rm S}$) color. A linear fit (solid black line) in each
    direction was used to model the correction. (\textit{Top-right}):
    VPD after the correction.}
  \label{fig5}
\end{figure}

We obtained astro-photometric catalogs for each image of each epoch as
described in Sect.~\ref{datared}, distortion corrected as described in
Sect.~\ref{GDC}. The VIRCAM photometry is calibrated by using stars in
common with the 2MASS catalog. We applied linear relations between the
VVV instrumental magnitudes and the 2MASS magnitudes based on
well-measured, unsaturated stars. We replaced the photometry of VVV
saturated stars with that of 2MASS.

The adopted reference frame is based on images taken on August
16$^{\rm th}$ 2012 (which have the best available seeing, are the
closest to the zenith, and are taken halfway between 2010 and
2014). The covered FoV is about
$1^\circ\!\!.1$$\times$$1^\circ\!\!.5$. We limited our PM analysis to
the innermost region of the field, within a radius of 20 arcmin from
the cluster center, where there is a significant number of cluster
members. We then computed the coefficients of the local
transformations to transform the stars' positions of each image into
the reference frame (\citealt{Ande06}). Local transformations reduce
most of the uncorrected distortion residuals and other systematic
effects that could harm our measurements. Indeed, the astrometric
accuracy reached in our reference master frame is $\sim$0.08 pixel (27
mas), more than three times larger than that described in
Sect.~\ref{GDC}. The main reason for this larger uncertainty is that
the VVV observations are not taken with an astrometric strategy in
mind (see discussion in Sect.~\ref{intro} and \ref{OBS}). Furthermore,
our \mbox{geometric-distortion} correction is an \textit{average}
solution, suited for $J$-filter images and at a specific epoch. The
fact that the positional residuals are three times larger also implies
that the distortion correction is not stable over time scales of 6
months. The local-transformation approach compensates for these
issues. 

In our local-transformation approach, we transformed the stellar
positions as measured in each image into the reference-frame system
using a subset of close-by, likely-cluster member (reference stars) to
a given star to compute its linear-transformation coefficients. As
such, our PMs are computed relative to the cluster mean motion, and
cluster members will end up around the (0,0) location on the
vector-point diagram (VPD). At the first iteration we selected the
reference stars for the local transformation based on their position
on the color-magnitude diagram (CMD). Once PMs were also estimated, we
improved our reference-star list by removing all those stars which
motion is not consistent with the cluster mean motion. \\

We computed the stellar displacements as the difference between the
transformed single-exposure positions and the master-frame
positions. In Fig.~\ref{fig3} we illustrate the \mbox{multi-epoch} PM
fit for a Bulge star (left panels) and for a cluster member (right
panels). In the top panels of the figure we show the stellar positions
transformed into the reference frame, \mbox{color-coded} according to
their epoch. In the bottom panels we show the displacements as a
function of time (relative to the reference epoch). The black lines
are the weighted least-squares fit to the data, where the weight is
defined as the square root of stars' \textsf{QFIT} (with
\mbox{poorly-measured} stars' having less weight). The proper motion
along the two directions ($\mu_x$, $\mu_y$) is the slope of the
straight lines. The relative PM errors are the formal errors of the
least-squares fit. The constant terms $x_0$ and $y_0$ indicate the
corrections to be applied to our reference-frame positions at epoch
J2012.62423. To exclude obvious outliers, for each star we iteratively
removed one point at a time from the sample, fit the proper motion
with the remaining points, and re-computed the vertical residual of
the removed point. We rejected all those points for which vertical
residuals were five times larger than the fit residual. A final fit to
all the remaining points provides our PM estimate for the star.

\begin{figure}
  \centering
  \includegraphics[width=\columnwidth]{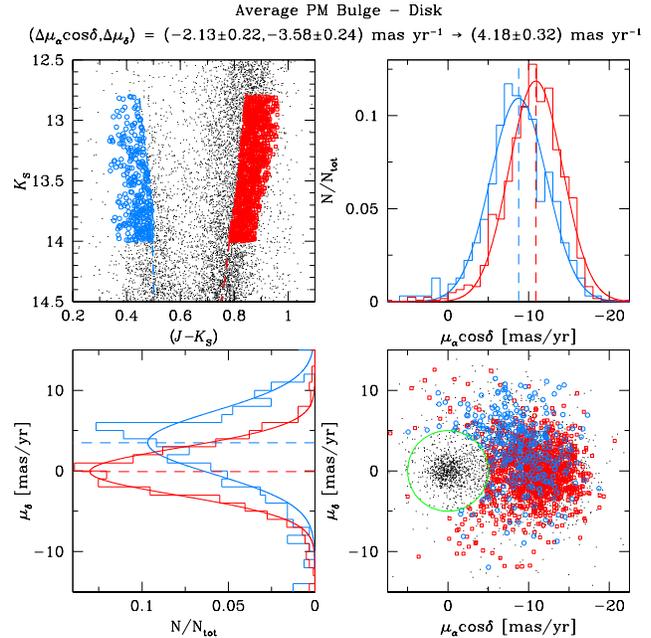}
  \caption{(\textit{Top-left}): $K_{\rm S}$ vs. ($J$$-$$K_{\rm S}$)
    CMD. We plotted Disk main sequence stars with blue open circles
    and Bulge red-giant-branch stars with red open squares. We limited
    our samples to 12.8$\le$$K_{\rm S}$$\le$14 in order to use only
    well-measured bright stars. We did not plot all stars within 5 mas
    yr$^{-1}$ from the center of the VPD to exclude most of the
    NGC~6656 members. (\textit{Bottom-right}): VPD with stars within
    12.8$\le$$K_{\rm S}$$\le$14. The green circle used to exclude
    cluster members has a radius of 5 mas
    yr$^{-1}$. (\textit{Bottom-left}): histograms of the $\mu_\delta$
    for the Bulge and Disk stars previously selected. We fitted each
    histogram with a single Gaussian. (\textit{Top-right}): as on
    \textit{Bottom-left} but for the $\mu_\alpha\cos\delta$.}
  \label{fig6}
\end{figure}

In Fig.~\ref{fig4} we show the $K_{\rm S}$ vs. ($J$$-$$K_{\rm S}$)
CMDs and the VPDs for the stars in the NGC~6656 field. In the VPDs we
show the ($\mu_\alpha\cos\delta$,$\mu_\delta$) PMs. In the left panel
we plotted the CMD of the entire sample of well-measured-PM
stars\footnote{We plot the PM errors as a function of the $K_{\rm S}$
  magnitude and drew by hand a fiducial line to remove obvious
  outliers. The cut is more important for faint stars, where PMs are
  less accurate. We used the same purging method for the stellar
  {\sf{QFIT}}. Furthermore, we kept only those stars measured in at
  least 50 exposures.} observed in the selected field, and split it
into nine bins, one $K_{\rm S}$-magnitude wide. In the middle-left
panels we show the corresponding VPD for each magnitude bin, in which
we drew a circle to enclose cluster-like-PM stars. The radius is a
compromise between including field stars with a cluster-like mean
motion and excluding cluster members with larger PM
uncertainties. Saturated and faint stars have poorly-measured PMs, so
the selection radii are more generous. In the middle-right panels we
show the histograms of the motion along the $\mu_\alpha\cos\delta$
direction (where the difference between cluster and field stars in the
PMs is more evident) to show the field contamination expected by our
selections. The rightmost panel shows the CMD of NGC~6656 members. The
horizontal branch, red-giant branch, main sequence turn-off and upper
main sequence are mostly cleaned by field stars and can be used for
studies of the properties of these stars in the near infrared. We also
analyzed the possible impact of differential-chromatic-refraction
effects on our PMs (Fig.~\ref{fig5}). We selected two samples of
NGC~6656 stars with a $K_{\rm S}$ magnitude between 13.5 and 15, one
on the horizontal branch (blue stars) and the other on the red giant
branch (red stars). Even if the mean motion of both samples is the
same within the error bars, we chose to remove the small contribution
of this effect (clearly visible in the $\mu_\alpha\cos\delta$
direction). All PMs in Fig.~\ref{fig4}, \ref{fig6} and \ref{fig7} have
been corrected accordingly. \\

To evaluate our PM precision we proceeded as follow. By construction,
our PMs are relative to the cluster mean motion. As such, the mean
location of cluster members on the VPD is (0,0), and the observed
dispersion of cluster members should in principle reflect the stellar
internal-motion dispersion plus our measurement errors. The internal
dispersion of NGC~6656 is about 0.5 mas yr$^{-1}$ (assuming a distance
of 3.2 kpc and a central velocity dispersion of 7.8 km s$^{-1}$,
\citealt{Har96}, 2010 edition). The 1-D dispersion (defined as the
68.27$^{\rm th}$ percentile of the distribution around the median) of
bright, unsaturated (13$\le$$K_{\rm S}$$\le$14) cluster stars in the
VPD is of about 1.5 mas yr$^{-1}$. By subtracting in quadrature the
internal dispersion of 0.5 mas yr$^{-1}$, we end up with an external
estimate of our PM precision, which is of about 1.4 mas yr$^{-1}$. \\

To further test of our astrometric accuracy we measured the relative
difference between the Bulge and the Disk bulk motion within the same
selected VVV field of NGC~6656 (Fig.~\ref{fig6}). To this aim, we
selected two samples of stars, one from the Disk main sequence and one
from the Bulge red giant branch. We considered only Disk (Bulge) stars
that in the $K_{\rm S}$ vs. ($J$$-$$K_{\rm S}$) CMD are bluer (redder)
than the respective fiducial line of the sequence. Furthermore, we
considered only those stars with PMs larger than 5 mas yr$^{-1}$ with
respect to the bulk motion of the cluster. We fit a single Gaussian to
the histograms of the Bulge and Disk PMs along each direction, and
found a relative displacement of
($\Delta\mu_\alpha\cos\delta$,$\Delta\mu_\delta$)$=$($-$2.13$\pm$0.22,$-$3.58$\pm$0.24)
mas yr$^{-1}$. The absolute difference between Bulge and Disk bulk
motions is therefore 4.18$\pm$0.32 mas yr$^{-1}$.

To test this result, we measured the relative displacement between the
Bulge and the Disk components using the motion of the same test stars
as measured in the UCAC4 and the PPMXL (\citealt{Roe10}) catalogs. We
found a relative displacement of 3.79$\pm$0.98 mas yr$^{-1}$ using
UCAC4 and 2.93$\pm$1.3 mas yr$^{-1}$ using PPMXL, which are in
agreement with our estimate within the error bars, though our estimate
has a smaller uncertainty. We also compared our measured difference
between Bulge and Disk motion with that predicted by the Besan\c{c}on
models (\citealt{Rob03}). We simulated both populations in the same
field covered by our application. We adopted an exponential trend for
photometric and PM errors as function of the magnitude to create a
model as close as possible to our data. The major challenge was to
take into account for the correct absorption toward the Galactic
plane. We used the Bulge Extinction And Metallicity (BEAM, see
\citealt{Gonz12,Gonz13}) calculator to compute the average extinction
in our field. This value, divided by NGC~6656 distance, gives us the
diffuse absorption of 0.15 mag kpc$^{-1}$. The difference between
Bulge and Disk motion obtained this way is 1.38$\pm$0.12 mas
yr$^{-1}$. This value is not consistent with our measurements. We
performed different simulations varying the absorption coefficients to
understand if the absorption law could somehow change the simulated
kinematics, but we found the results were about the same. We attribute
this significant difference to the difficulty of the Besan\c{c}on
model in simulating the reddening, Galaxy stellar densities and
kinematics toward the Galactic Plane where the extinction is high. \\

\subsection{Future perspectives}

\begin{figure}
  \centering
  \includegraphics[width=\columnwidth]{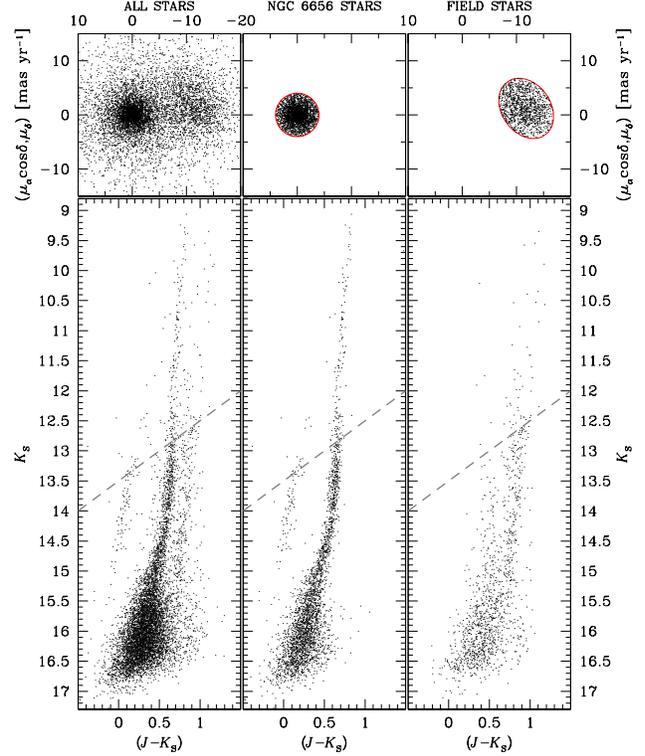}
  \caption{(\textit{Top}): vector-point diagrams with a time baseline
    of about six years obtained combining the HAWK-I and VVV
    data. (\textit{Bottom}): $K_{\rm S}$ vs. ($J$$-$$K_{\rm S}$)
    color-magnitude diagrams of the stars in common between the HAWK-I
    and VVV fields. We plot only well-measured-PM stars. On the left
    panels we plot the entire sample, while in the center and right
    panels we plot only cluster members and field stars respectively.
    We considered NGC~6656 stars those stars with a proper motion
    within 4 mas yr$^{-1}$ around the cluster mean motion (red circle
    centered in (0,0) in the middle VPD); while the stars enclosed in
    the ellipse centered in ($-$11.5,1.2) mas yr$^{-1}$ with major and
    minor axes of 12 and 9 mas yr$^{-1}$ are probable field stars.}
  \label{fig7}
\end{figure}

The VISTA Variables in the V\'ia L\'actea will be completed in 2016
\citep{He14}, and the time baseline provided by the uniform VVV data
will be about six years. As an example, we combined the VVV images of
NGC~6656 and the HAWK-I data previously used by \cite{Libra14}. Since
the HAWK-I images were taken in 2007, we used only the VVV archival
images between 2010 and 2013 in order to have approximately the 6
years of time baseline. We computed the PMs as described in the
previous section and in Fig.~\ref{fig7} we show the resulting CMDs and
VPDs. As expected, with a larger time baseline we are able to
completely separate cluster and field stars. This example shows again
the great astrometric potential of the full-baseline VVV data. Older
epochs (both optical and near-infrared data) are available in the
archives, and the proper motions will be an invaluable resource to
distinguish the different stellar populations in the Galaxy.

\section{Conclusions}

In this paper we present our reduction pipeline for the VIRCAM
detector and the geometric distortion solution based on the $J$
filter. Thanks to our distortion correction and to the adopted
dithered observing strategy, we are able to reach a positional
residual of $\sim$8 mas in each coordinate in each exposure across the
entire FoV of VIRCAM. Note that we are talking about relative
astrometry. Our absolute astrometry is not as good as the relative one
because the linear terms are constrained only with 2MASS.

We release a \texttt{FORTRAN} routine to correct the geometric
distortion. For a given position in a single-chip raw frame ($x^{\rm
  raw}$,$y^{\rm raw}$) and the chip number, the code produces ($x^{\rm
  corr}$,$y^{\rm corr}$) coordinates in the meta-frame reference
system. The code is available at our group's web
page\footnote{\href{http://groups.dfa.unipd.it/ESPG/}{http://groups.dfa.unipd.it/ESPG/}}. The
use of this distortion solution is encouraged regardless of the
specific method adopted to measure stellar positions. Each meta
catalog is projected into a plane tangential to its center. This
offers the best single-catalog, distortion-free positions. Please note
that, in order to construct a common reference frame, all meta
catalogs should be instead projected into the same tangent plane (see
Sect.~\ref{GDC}).

As a test bed of the astrometric accuracy reached by our
geometric-distortion correction, we applied our reduction pipeline to
a set of VVV archival images. We chose a field centered on the
globular cluster NGC~6656 and we computed the relative proper motion
of the NGC~6656 and Galactic bulge and disk stars, as well as the
individual motion of each star in the field. We noticed that our
astrometric accuracy is worse ($\sim$0.08 pixel) using VVV data. Our
\mbox{geometric-distortion} correction is an average solution and the
distortion is not entirely stable. However, by starting with a good
average solution, local transformations (used to compute the proper
motions) can be used to efficiently achieve optimal precision even
with this type of data. We demonstrate that we are able to separate
cluster and background/foreground field stars with a time baseline of
only four years. The cluster stars, in the cleaned CMD, can be used
for the study of the stellar populations of NGC~6656.  We also showed
that the field stars, in the direction of NGC~6656, are of great use,
e.g., to separate (and study) the proper motions of the Galactic disk
and bulge components. We demonstrated that our results are consistent
with what can be obtained using UCAC4 and PPMXL catalogs, though our
measurements have a much smaller error. Galactic models fail to
reproduce the observations, likely because of the difficulties to
reproduce the reddening and kinematics towards the Galactic bulge.

With the images analyzed in this paper and a time baseline of about
four years, we obtained a typical astrometric precision of 1.4 mas
yr$^{-1}$ for bright, unsaturated well-measured stars. This value
corresponds to $\sigma_v$$\sim$21 km s$^{-1}$ at the distance of
NGC~6656 (3.2 kpc from \citealt{Har96}, 2010 edition), or $\sim$53 km
s$^{-1}$ at 8 kpc (a reference distance for the Bulge). At the end of
the VVV survey, the total time baseline will be of about six years,
thus further increasing the final achievable PM accuracy. The use of
older, archive, optical and near-infrared data will further enhance
the proper-motion capability of the VVV survey. The astrometric
capability of this survey is complementary to GAIA, in particular in
the most crowded and heavily-absorbed regions not reachable by GAIA,
and to study objects below its magnitude limit ($G$$\sim$20).

\section*{Acknowledgements}

M.L. and G.P. acknowledge partial support from the Universit\`a degli
Studi di Padova CPDA101477 grant. M.L. acknowledges support from the
STScI 2013 DDRF Spring program ``High-precision astrometry with
wide-field detectors'' (PI: Bellini). V.N. acknowledges support from
``Studio preparatorio per le osservazioni della missione ESA/CHEOPS''
(\#42/2013). M.Z. acknowledges funding from the BASAL Center for
Astrophysics and Associated Technologies PFB-06, Proyecto FONDECYT
Regular 1110393, and the Ministry for the Economy, Development, and
Tourism's Programa Iniciativa Científica Milenio through grant
IC120009, awarded to the Millennium Institute of Astrophysics
(MAS). We thank the anonymous referee for the useful comments and
suggestions that considerably improved the quality of our paper.

\appendix

\section{Geometric distortion maps}
\label{appendix}

\begin{figure*}
  \centering
  \includegraphics[width=\columnwidth]{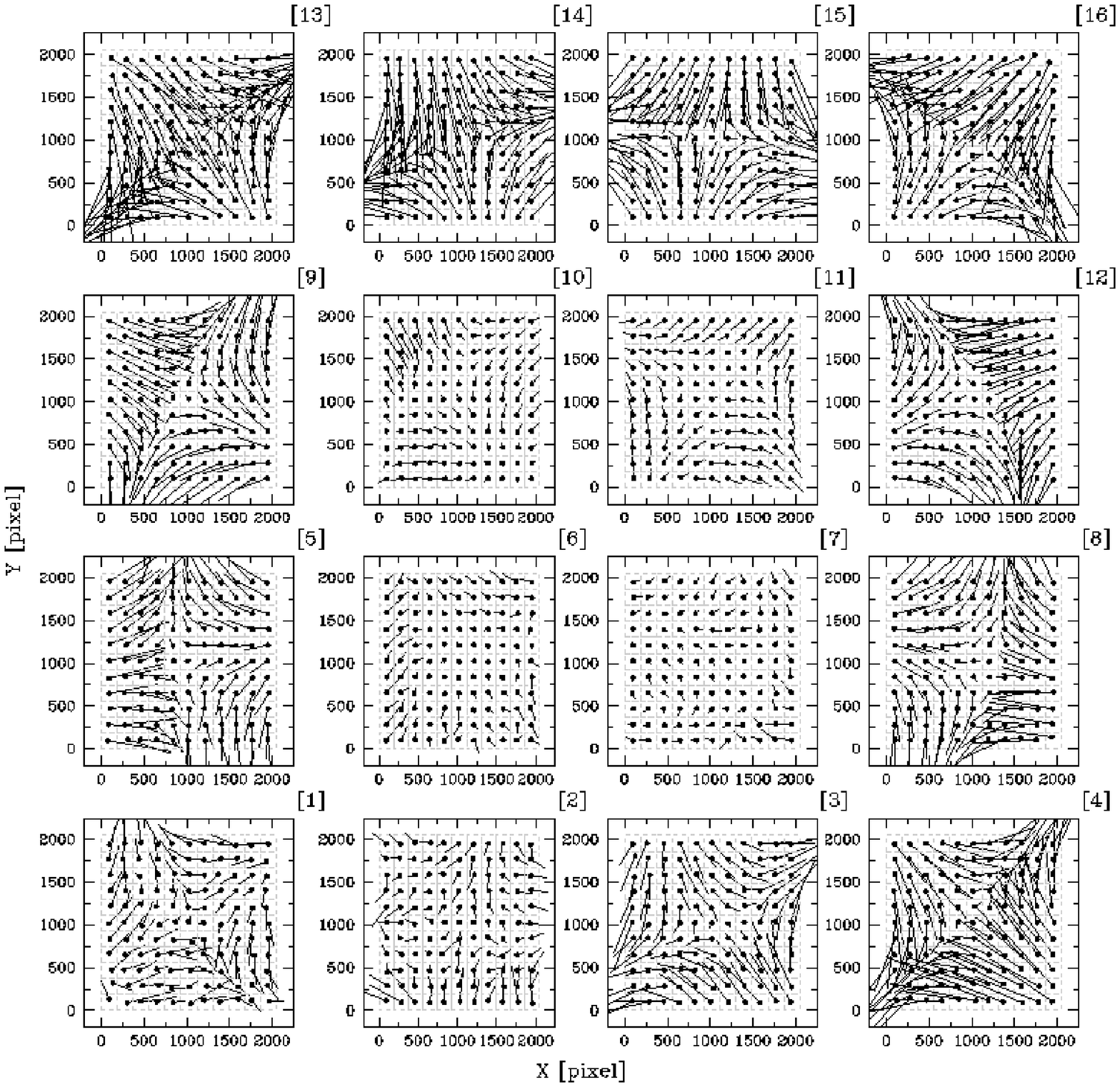}
  \hspace{0.5 cm}
  \includegraphics[width=\columnwidth]{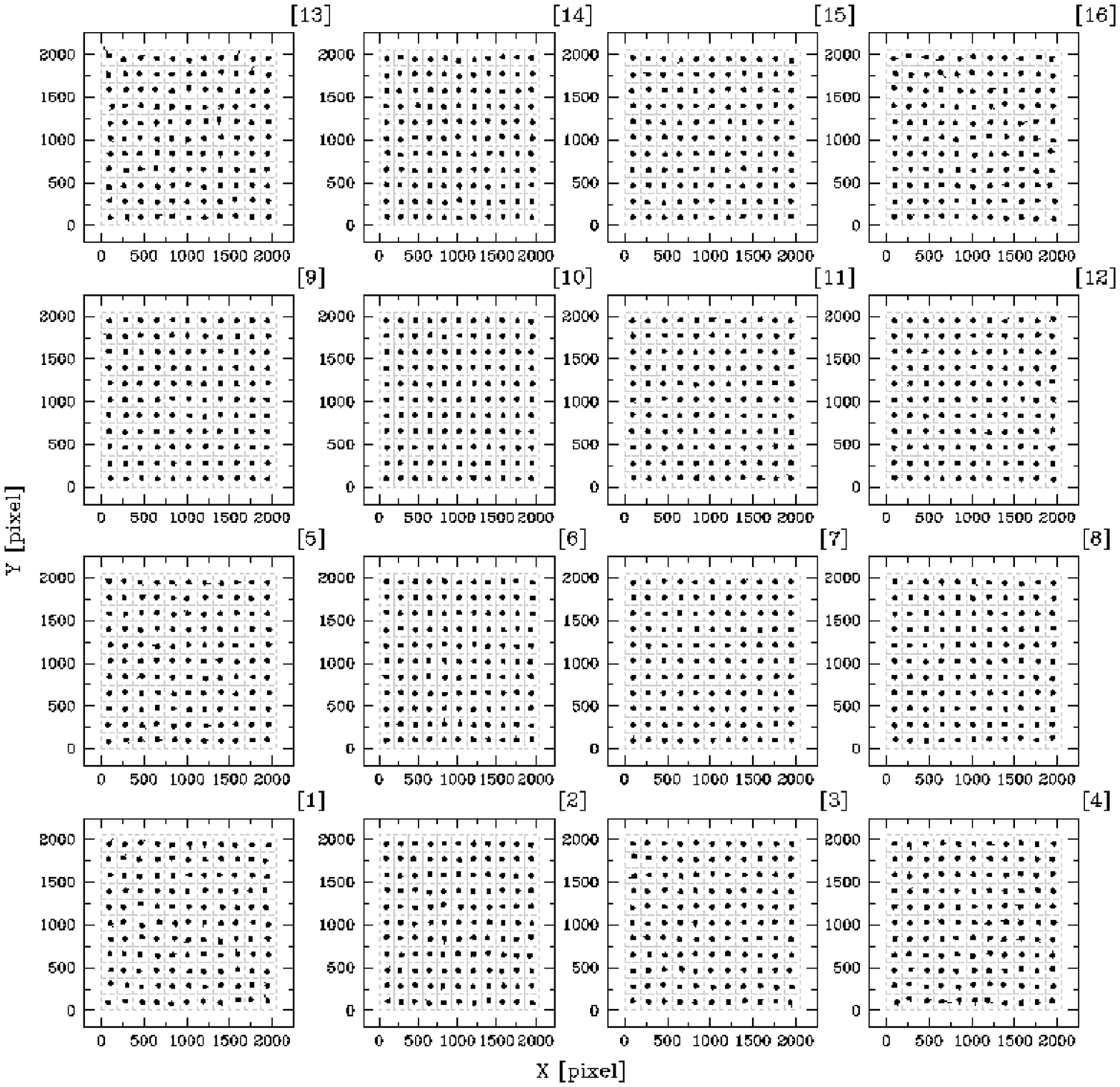}
  \caption{(\textit{Left}): Residual trends for the 16 chips when we
    use uncorrected stellar positions. The labels on the top-right
    corner of each box represent the chip number. The size of the
    residual vectors is magnified by a factor of 250. Some degree of
    distortion is clearly visible in the outermost
    chips. (\textit{Right}): Residuals after our distortion correction
    is applied. The size of the residual vectors is now magnified by a
    factor 5000.}
  \label{app1}
\end{figure*}

\begin{figure*}
  \centering
  \includegraphics[width=\columnwidth]{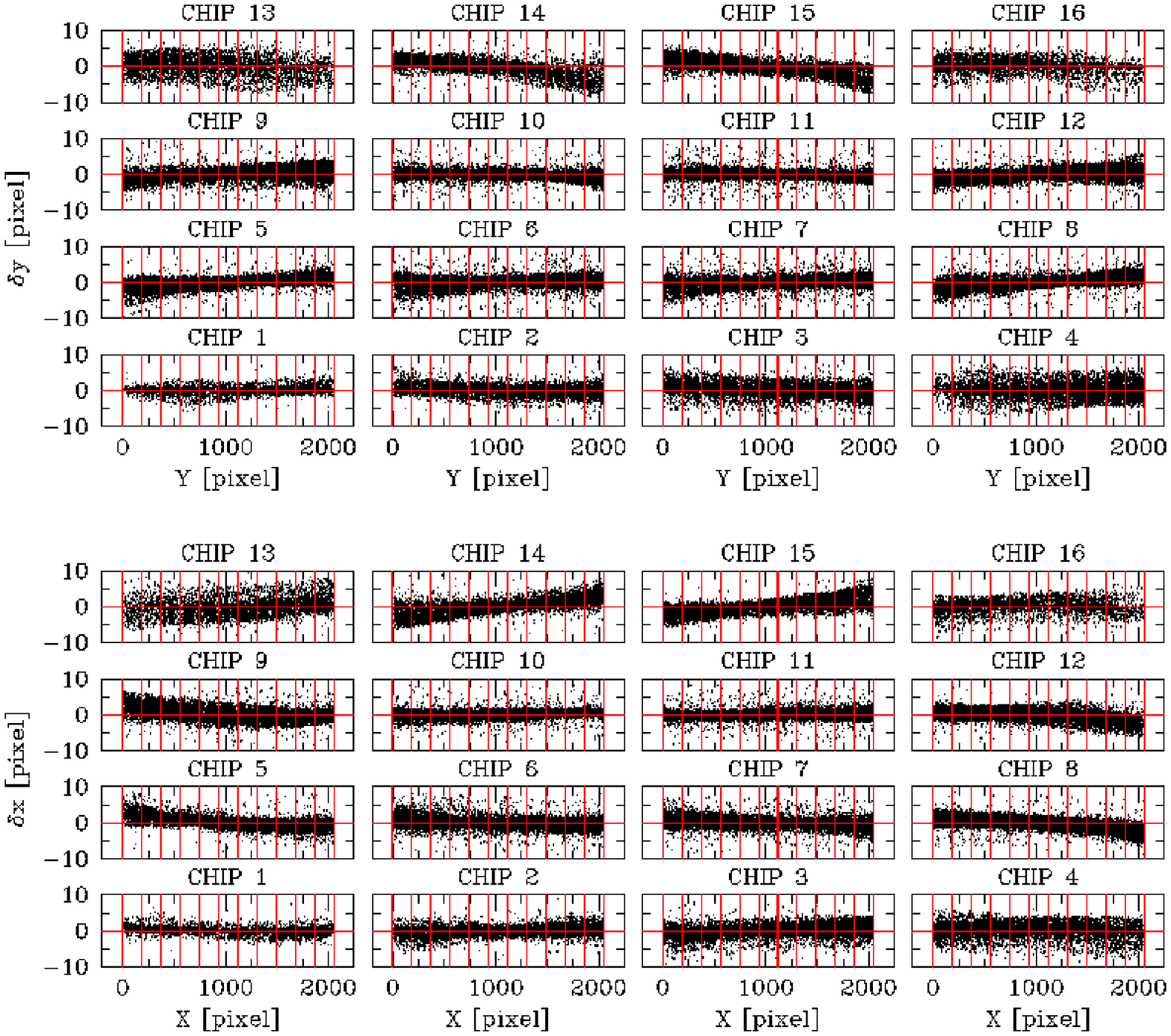}
  \hspace{0.5 cm}
  \includegraphics[width=\columnwidth]{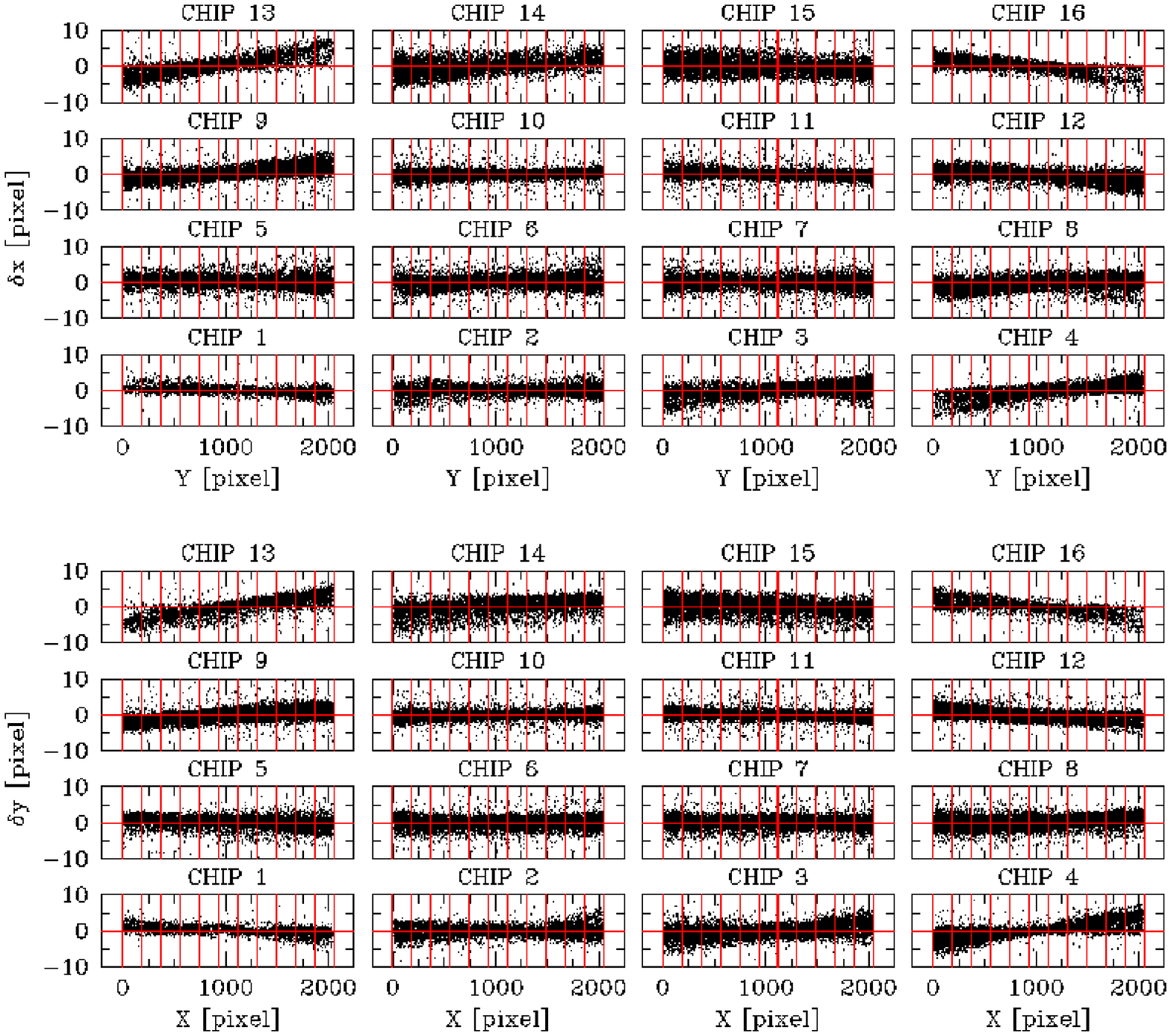}
  \caption{From the \textit{Bottom-left} panels, clockwise: $\delta x$
    vs. $X$, $\delta y$ vs. $Y$, $\delta x$ vs. $Y$ and $\delta y$
    vs. $X$ for each of the 16 VIRCAM chips before we applied the
    distortion correction.}
  \label{app2}
\end{figure*}

\begin{figure*}
  \centering
  \includegraphics[width=\columnwidth]{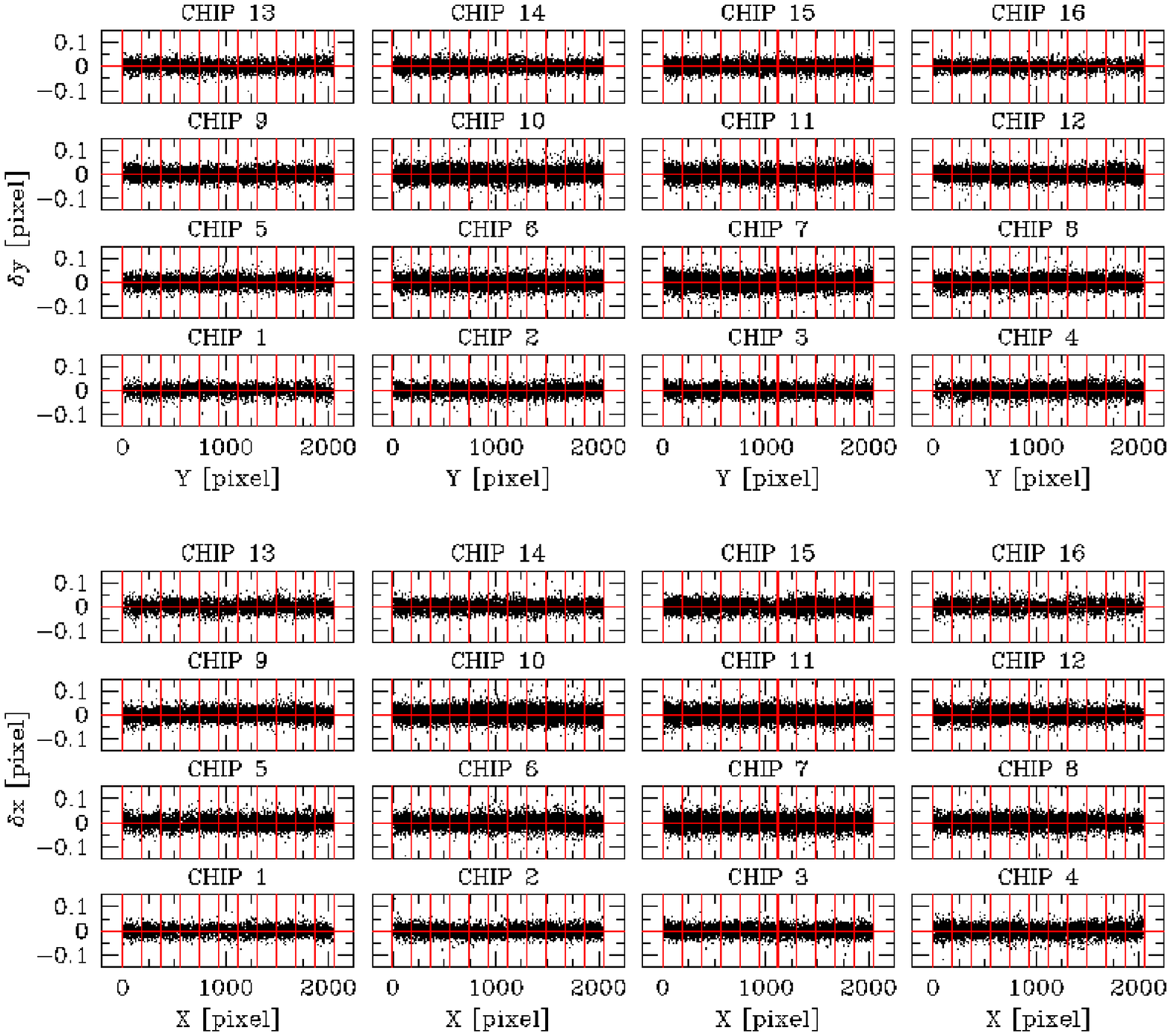}
  \hspace{0.5 cm}
  \includegraphics[width=\columnwidth]{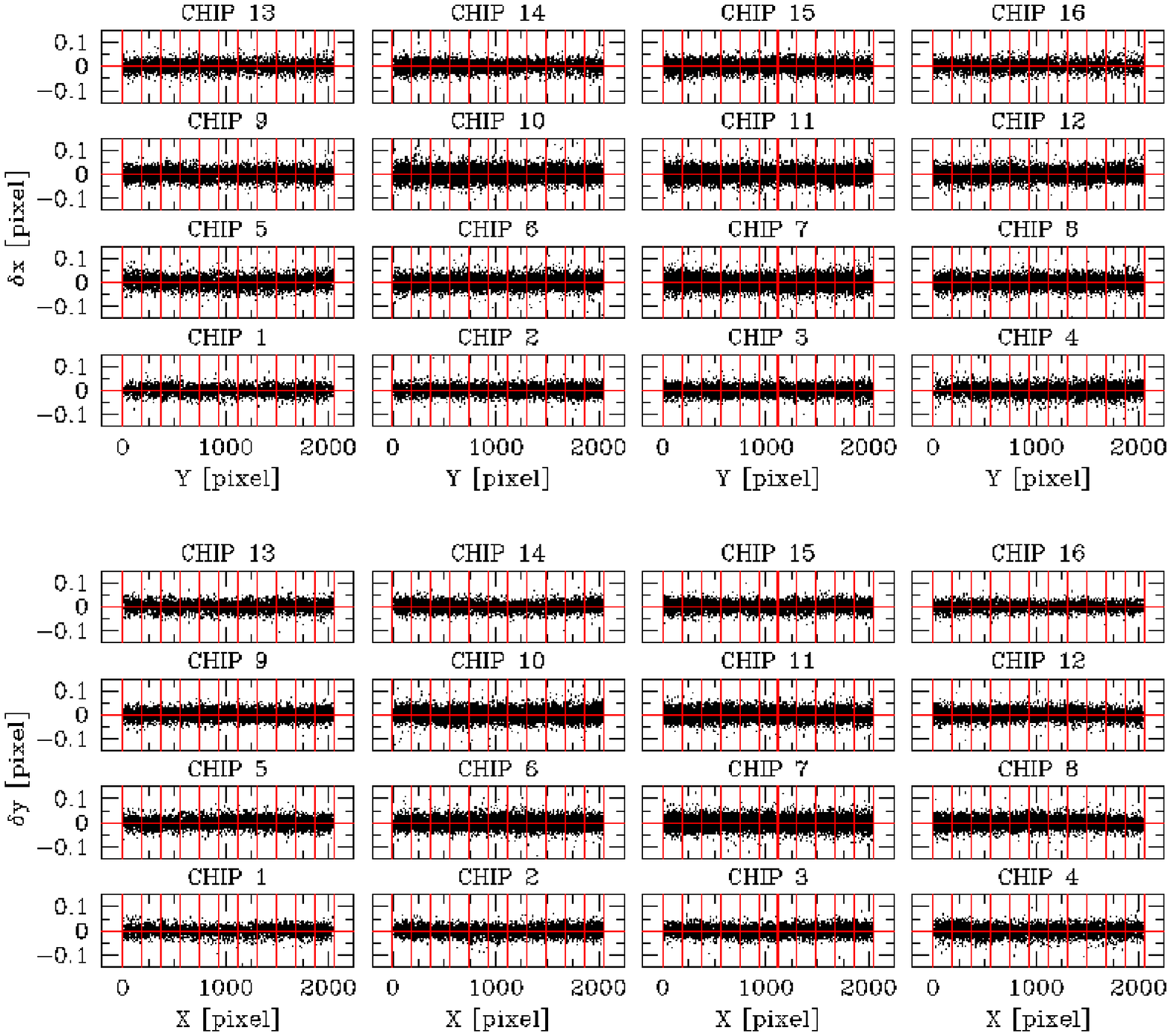}
  \caption{As in Fig.~\ref{app2} but after the distortion correction
    is applied.}
  \label{app3}
\end{figure*}

In this appendix we show the distortion maps (Fig.~\ref{app1}) and the
positional residuals (Fig.~\ref{app2} and Fig.~\ref{app3}) for the 16
chips of VIRCAM before and after we applied the distortion solution
described in Sect.~\ref{GDC}.

\label{lastpage}

\end{document}